\documentclass{amsart}
\usepackage[centertags]{amsmath}
\usepackage{amsfonts,amssymb}
\usepackage{enumerate}
\usepackage{amsfonts,amsmath,amssymb,cite,graphicx,color,ulem,amsthm}
\usepackage{fourier}
\usepackage[margin=1.5in]{geometry}
\usepackage{hyperref}

\def\RR{\mathbb{R}}

\renewcommand{\d}[2]{\frac{d #1}{d #2}} % for derivatives
\newcommand{\avg}[1]{\langle #1 \rangle}

\newcommand{\avgg}[1]{\left\langle #1 \right\rangle}

\newcommand{\grad}{\nabla}
\renewcommand{\div}{\nabla\cdot}

\def\bb{\boldsymbol{b}}

\def\qb{\boldsymbol{q}}

\def\pb{\boldsymbol{p}}

\def\xb{\boldsymbol{x}}

\def\Xb{\boldsymbol{X}}

\def\thetab{\boldsymbol{\theta}}

\theoremstyle{plain}

\theoremstyle{remark}

\def\<{\langle} \def\>{\rangle}

\title[NONEQUILIBRIUM IMPORTANCE SAMPLING]{DYNAMICAL COMPUTATION OF THE
  DENSITY OF STATES AND BAYES FACTORS\\ USING NONEQUILIBRIUM IMPORTANCE
  SAMPLING}

\author{Grant Rotskoff$^{1,2}$} \author{Eric Vanden-Eijnden$^1$}
\address{$^1$Courant Institute of Mathematical Sciences, New York
  University, 251 Mercer Street, New York, NY 10012}  \address{$^2$Present
  address: Department of Chemistry, Stanford University, 165 Keck
  Science Building, Stanford, CA 94305}

\begin{document}
\numberwithin{equation}{section}

\begin{abstract}
  Nonequilibrium sampling is potentially much more versatile than its
  equilibrium counterpart, but it comes with challenges because the
  invariant distribution is not typically known when the dynamics
  breaks detailed balance.  Here, we derive a generic importance
  sampling technique that leverages the statistical power of
  configurations transported by nonequilibrium trajectories, and can
  be used to compute averages with respect to arbitrary target
  distributions.  As a dissipative reweighting scheme, the method can
  be viewed in relation to the annealed importance sampling (AIS)
  method and the related Jarzynski equality.  Unlike AIS, our approach
  gives an unbiased estimator, with provably lower variance than
  directly estimating the average of an observable.  We also establish
  a direct relation between a dynamical quantity, the dissipation, and
  the volume of phase space, from which we can compute quantities such
  as the density of states and Bayes factors.  We illustrate the
  properties of estimators relying on this sampling technique in the
  context of density of state calculations, showing that it scales
  favorable with dimensionality---in particular, we show that it can
  be used to compute the phase diagram of the mean-field Ising model
  from a single nonequilibrium trajectory.  We also demonstrate the
  robustness and efficiency of the approach with an application to a
  Bayesian model comparison problem of the type encountered in
  astrophysics and machine learning.
\end{abstract}

%\date{\today}


\maketitle

\section{Introduction}
\label{se:intro}

Statistical estimation using averages over a dynamical process
typically relies on the principle of detailed balance.  Consider a
dynamical system
\begin{equation}
 \dot{\Xb}(t,\xb) = \bb({\Xb}(t,\xb)) \qquad \Xb(0,\xb) = \xb
 \label{eq:dynamics}
\end{equation}
where $\xb \in \Omega \subset \mathbb{R}^n$ is a state that is
propagated in time to ${\Xb}(t,\xb)$ via the vector field
$\bb: \Omega \to \mathbb{R}^n$.  If the dynamics is microscopically
reversible with respect to some target density $\rho(\xb)$, then this
density is preserved under time evolution.  Practically, this means
that the expectation of an observable $\phi(\xb)$ with respect to
$\rho(\xb)$, which we denote by $\langle \phi\rangle$, can be computed
as a time average along an equilibrium trajectory generated
from~\eqref{eq:dynamics}, provided that the dynamics is ergodic.  This
direct sampling scheme becomes inefficient if the expectation
$\langle \phi\rangle$ is dominated by values of $\xb$ that are rare
under $\rho(\xb)$ and therefore infrequently visited by the
dynamics~\eqref{eq:dynamics}.

Importance sampling estimates relying on nonequilibrium dynamics have
shown success in a variety of applications, from statistical physics
to machine
learning~\cite{jarzynski_nonequilibrium_1997,hummer_free_2001,%
  sun_equilibrium_2003,athenes_path-sampling_2004,
  dellago_computing_2013,neal_annealed_2001,nilmeier_2011gr}.  Here,
we derive a class of estimators based on an exact reweighting of the
samples gathered during a nonequilibrium process with a stationary
density.  Similar to the annealed importance sampling (AIS)
method~\cite{neal_annealed_2001} and estimators based on the Jarzynski
equality~\cite{jarzynski_nonequilibrium_1997}, our scheme accelerates
the transport of density to rare regions of phase space which may make
substantial contributions to equilibrium averages.  This basic idea is
exploited by many different enhanced sampling
techniques~\cite{allen_forward_2009,%
  bouchard-cote_bouncy_2018,peters_rejection-free_2012,michel_generalized_2014}.
Physically, the statistical weight of the transported density can be
interpreted through the fluctuation theorem as a dissipative
reweighting whose value can be derived explicitly.  As we show, the
resulting estimator is unbiased, unlike the AIS estimator, which
requires computing a ratio of sample means (cf. Eq. (12) of
Ref.~\cite{neal_annealed_2001}).  Our estimator always has lower
variance than the direct estimator, a reduction that comes at the
nontrivial cost of generating trajectories.  That said, nonequilibrium
transport enables us to access states that are exponentially rare in
original density but which may dominate expectation values; direct
sampling fails dramatically in such settings.

\section{Nonequilibrium estimators}
\label{sec:neet}

A generic importance sampling
scheme to compute the average with respect to some target density
$\rho(\xb)$ reweights samples drawn from another density
$\rho_{\text{ne}} (\xb)$
\begin{equation}
  \label{eq:reweight}
  \avg{\phi} = \avg{ \phi \rho/ \rho_{\text{ne}}}_{\text{ne}}.
\end{equation}
Our samplers use for $\rho_{\text{ne}}(\xb)$ the non-equilibrium
stationary density of a dynamical system based on generating
trajectories by an initiate-then-propagate procedure: We draw points
$\xb$ from the density $\rho(\xb)$ that we then propagate forward and
backward in time using the dynamics~\eqref{eq:dynamics} until the
trajectories $\Xb(t,\xb)$ hit some fixed target set.  Concrete
applications determine the appropriate choice of target sets: they
could for example be the boundary of $\Omega$ or the fixed points
of~\eqref{eq:dynamics} in $\Omega$~\footnote{ As made clear by the
  calculation in the supplementary material, the main requirement to
  derive the estimator \eqref{eq:estimator} is that $\tau^\pm(\xb)$
  satisfy $\tau^\pm(\Xb(t,\xb))= \tau^\pm(\xb)-t$ for all
  $t\in[\tau^-(\xb),\tau^+(\xb)]$. This requirement is satisfied with
  $\tau^\pm(\xb)$ defined as the times when the trajectory
  $\Xb(t,\xb)$ hits fixed target sets.  }.  Using the set of
trajectories generated this way we define the nonequilibrium average
$\avg{\cdot}_{\text{ne}} $ as
\begin{equation}
  \begin{aligned}
  \label{eq:rrhoneq}
  \avg{\phi}_{\text{ne}} &= \frac1{\avg{\tau} } \int_{\Omega}
  \int_{\tau^-(\xb)}^{\tau^+(\xb)} \phi(\Xb(t,\xb)) dt \, \rho(\xb)
  d\xb  % \\
  % &\equiv \int_{\Omega} \phi(\xb) \rho_{\text{ne}} (\xb) d\xb
\end{aligned}
\end{equation}
where $\tau^+(\xb)\ge0$ and $\tau^-(\xb)\le0$ are the first times at
which $\Xb(t,\xb)\in \partial\Omega$ in the future or in the past,
respectively, and $\avg{\tau}=\avg{\tau^+} -\avg{\tau^-}$. By changing integration variables using
$X(t,\xb)\to \xb$ and $t\to-t$ we can express~\eqref{eq:rrhoneq} as
\begin{equation}
  \begin{aligned}
  \label{eq:rrhoneq2}
  \avg{\phi}_{\text{ne}} = \frac1{\avg{\tau} } \int_{\Omega} \phi(\xb)
  \int_{\tau^-(\xb)}^{\tau^+(\xb)} \!\!\!J(t,\xb)
  \rho(\Xb(t,\xb)) dt \, d\xb
\end{aligned}
\end{equation}
where $J(t,\xb)$ is the Jacobian of the transformation:
\begin{equation}
  \label{eq:jacobian}
  J(t,\xb) = \exp \left({\textstyle\int_0^t \div \bb (\Xb(s, \xb))
      ds} \right).
\end{equation}
Physically, the Jacobian corresponds to the total energy dissipation
up to time $t$ along the trajectory.  This derivation is described in
detail in Appendix~\ref{sec:derirhone}).  Now,~\eqref{eq:rrhoneq2} can
be interpreted as an expectation with respect to a nonequilibrium
density,
$\avg{\phi}_{\text{ne}}=\int_{\Omega} \phi(\xb) \rho_{\text{ne}} (\xb)
d\xb$, with $\rho_{\text{ne}} (\xb)$ given by
\begin{equation}
  \label{eq:rrhoneqexplicit}
  \begin{aligned}
    \rho_{\text{ne}}(\xb) = \frac1{\avg{\tau}}
    \int_{\tau^-(\xb)}^{\tau^{+}(\xb)}J(t,\xb)\rho(\Xb(t,\xb))
    dt.
  \end{aligned}
\end{equation}
We can now use this expression for $\rho_{\text{ne}}(\xb)$
in~\eqref{eq:reweight} for reweighting.  As shown in
Appendix~\ref{Sec:derivest1} this gives~\footnote{%
  Note that, while \eqref{eq:rrhoneqexplicit} requires that
  ${\avg{\tau}}$ be finite, \eqref{eq:estimator0} does not, and this
  equation holds as long as long as the integrals in it converge. In
  fact it is easy to prove~\eqref{eq:estimator0} directly by
  performing a few changes of integration variable, as shown in
  Appendix~\ref{Sec:derivest1} and~\ref{sec:fixedest}.}
\begin{equation}
  \begin{aligned}
    \avg{\phi} =
    \avgg{\frac{\int_{\tau^-}^{\tau^+}\phi(\Xb(t))J(t)\rho(\Xb(t))
        dt}{\int_{\tau^-}^{\tau^+}J(t)\rho(\Xb(t)) dt} }
  \end{aligned}
  \label{eq:estimator0}
\end{equation}
which yields the estimator, one of our main results,
\begin{equation}
  \avg{\phi} = \lim_{N\to\infty} \avg{\phi}_N\quad
  \text{where} \quad \avg{\phi}_N = \frac{1}{N} \sum_{i=1}^N
  \frac{\int_{\tau^-(\xb_i)}^{\tau^+(\xb_i)}\phi(\Xb(t,\xb_i))J(t,\xb_i)\rho(\Xb(t,\xb_i))
    dt}{\int_{\tau^-(\xb_i)}^{\tau^+(\xb_i)}
    J(t,\xb_i)\rho(\Xb(t,\xb_i)) dt},
  \label{eq:estimator}
\end{equation}
provided that the points $\xb_i$ are drawn (not necessarily
independently) from $\rho(\xb)$. An equivalent
to~\eqref{eq:estimator0} using no target set and running the
trajectories on a finite time lag instead is given in
Appendix~\ref{sec:fixedest}. The estimator
\eqref{eq:estimator0}reweights points sampled along a nonequilibrium
trajectory according to their dissipation, a physically analogous
strategy to that in AIS (see Appendix~\ref{sec:AIScomp} for a more
detailed comparison with AIS).

Unlike other dissipative reweighting strategies, the estimator
$\avg{\phi}_N $ is unbiased, valid for any
dynamics~\eqref{eq:dynamics} and any target density $\rho(\xb)$.  Like
standard Metropolis Monte-Carlo, it only requires knowing the density
up to a normalization factor.  It has lower variance than the direct
estimator $N^{-1}\sum_{i=1}^N \phi(\xb_i)$ ~\footnote{%
  Note that this comparison neglects the cost of generating the ascent
  / descent trajectories from the points $\xb_i$.}  since the variance
of this direct estimator is $N^{-1}(\avg{\phi^2}-\avg{\phi}^2)$
whereas the variance of $\avg{\phi}_N$ is $N^{-1}(A-\avg{\phi}^2)$
with $A\le \avg{\phi^2}$ since Jensen's inequality implies
\begin{equation}
  \label{eq:3}
    A = \avgg{\Bigg|\frac{\int_{\tau^-}^{\tau^+}\phi(\Xb(t))
        J(t)\rho(\Xb(t)) dt}{\int_{\tau^-}^{\tau^+}J(t)\rho(\Xb(t))
        dt}
      \Bigg|^2} 
    \le \avgg{\frac{\int_{\tau^-}^{\tau^+}|\phi(\Xb(t))|^2J(t)\rho(\Xb(t))
        dt}{\int_{\tau^-}^{\tau^+}J(t)\rho(\Xb(t)) dt}} =
    \avgg{\phi^2}.
\end{equation}
With a proper choice of $\bb(\xb)$ the estimator $\avg{\phi}_N $ in
\eqref{eq:estimator} has the potential to significantly outperform the
direct estimator (in Appendix~\ref{sec:zerovar} we show that
$\avg{\phi}_N $ is a zro-variance estimator in one-dimension). It does
so by transporting points drawn naively from $\rho(\xb)$ towards
regions that statistically dominate the expectation of $\phi(\xb)$. In
practice, it is also simple to use: (i) sample points $\xb_i$ from
$\rho(\xb)$ using e.g. standard Monte Carlo; (ii) compute the
trajectory $\Xb(t,\xb_i)$ passing through each of these points by
integrating~\eqref{eq:dynamics} forward and backward in time until
$\Xb(\tau^\pm (\xb_i),\xb_i) \in \partial \Omega$ (which also gives
$\tau^\pm (\xb_i)$); (iii) use these data to calculate $J(t,\xb_i)$
from~\eqref{eq:jacobian} first, then the integrals
in~\eqref{eq:estimator}; (iv) average the results to get
$\avg{\phi}_N$. Note that the operations in (ii) and (iii) can be
performed in parallel, and we can monitor the value of the running
average $\avg{\phi}_N$ as $N$ increases to check convergence.

Let us also note that a discrete-time equivalent of
\ref{eq:estimator0} has now been given in~\cite{thin2021invertible}.

\section{Density of states (DOS)}
\label{sec:dos}

Consider a $d$-dimensional system
with position $\qb\in \mathbb{R}^d$, momenta $\pb\in \mathbb{R}^d$,
and Hamiltonian
\begin{math}
  \label{eq:1}
  \mathcal{H}(\qb,\pb) =  \tfrac12 |\pb|^2 + U(\qb)
\end{math}
where $U(\qb)$ is some potential bounded from below. Let $V(E)$ be the
volume of phase space below some threshold energy $E$,
\begin{equation}
V(E) = \int_{\mathcal{H}(\qb,\pb) < E}\  d\qb d\pb,
\label{eq:vos}
\end{equation}
From $V(E)$ one can compute the DOS, $D(E)=V'(E)$, or the
canonical partition function,
$Z(\beta) = \int_{\mathbb{R}} e^{-\beta E} D(E) dE= \beta
\int_{\mathbb{R}} e^{-\beta E} V(E) dE$.

To calculate~\eqref{eq:vos} with our estimator~\eqref{eq:estimator},
we set $\xb=(\qb,\pb)$, define
$\Omega = \{(\qb,\pb): \mathcal{H}(\qb,\pb) < E_{\max}\}$ for some
$E_{\max} < \infty$, and use dissipative Langevin dynamics with
$\bb(\xb) = (\pb , -\grad U(\qb) - \gamma \pb)$ in~\eqref{eq:dynamics}
\begin{equation}
  \dot{\qb} = \pb, \qquad
  \dot{\pb} = -\grad U(\qb) - \gamma \pb,
\label{eq:langevin}
\end{equation}
for some friction coefficient $\gamma>0$.  With this choice, the
dissipative term in the estimator~\eqref{eq:estimator} takes the
simple form:
\begin{equation}
J(t,\xb) =e^{-d\gamma t}.
\end{equation}
If we also choose the target density $\rho(\xb)$ to be uniform in
$\Omega$, the estimator further simplifies due to cancelation of the
two $\rho$ terms in~\eqref{eq:estimator}. We arrive at
\begin{equation}
  \begin{aligned}
  \frac{V(E)}{V(E_{\text{max}})} =\lim_{N\to\infty} \frac{1}{N}
  \sum_{i=1}^N
  e^{-d\gamma \left(\tau^E(\xb_i)-\tau^-(\xb_i)\right)}
  \label{eq:dos}
\end{aligned}
\end{equation}
where $\tau^E(\xb)$ denotes the positive (and possibly infinite) or
negative time for a trajectory initiated from $\xb=(\qb,\pb)$ to reach
energy $E\le E_{\text{max}}$ under the dynamics~\eqref{eq:langevin}.
Eq.~\eqref{eq:dos} is our second main result: it establishes a
dictionary between a nonequilibrium dynamical quantity and a purely
static, global property of the energy landscape, $V(E)$.
This result asserts that the rate of decrease of the volume of phase space can be measured by computing an average of the total dissipation of nonequilibrium descent trajectories.
We do not know of an analogous result in the literature.

The $\tau^+(\xb)$ terms vanish in this dynamics because the time to reach
a local minimum diverges \footnote{%
  This indicates that the density $\rho_{\text{ne}}(\xb)$ is singular
  here because the density becomes atomic on the local minima of
  $\mathcal{H}(\xb)$. However the estimator~\eqref{eq:estimator0}
  remains valid and explicitly given by~\eqref{eq:dos}.}.  In practice
we halt the forward trajectories when the norm of the gradient is
below some tolerance.  To compute an unnormalized volume, we can
estimate $V(E_{\text{max}})$ with standard Monte Carlo integration.

The power of the procedure we have described comes from the fact that
the forward trajectories are guaranteed to visit regions of low energy
around local minima of $U(\qb)$ that would otherwise be difficult to
sample by drawing points uniformly in
$\{\mathcal{H}(\qb,\pb) < E_{\max}\}$.  In this regard our approach is
also similar to nested sampling~\cite{skilling_nested_2006,
  brewer_diffusive_2011, partay_nested_2014,
  martiniani_superposition_2014, bolhuis_nested_2018}. Like nested
sampling, we do not require an
\textit{a~priori} stratification of the energy shells, which is the
way the DOS is typically calculated via thermodynamic
integration~\cite{frenkel_understanding_2001,gelman_simulating_1998} or simulated
tempering~\cite{marinari_simulated_1992,earl_parallel_2005}.
% formula~\cite{dinner_2018fj}
Our method also offers several
advantages over nested sampling.  First, the depth of energies reached
in nested sampling is determined by the initial number of points used
in a computation.  If too few points are used, the calculation must be
repeated in full with a larger number of initial points.  Here, the accuracy of the calculation improves and explores deeper minima
simply by running additional ascent/descent trajectories.  In
addition, our approach does not require uniform sampling below
\textit{every} energy level, which is required in nested sampling and is
a difficult condition to
implement~\cite{martiniani_superposition_2014}.  We must only generate
points uniformly below the highest energy level, $E_{\max}$, which is
usually much easier.  Computationally, we also benefit from the fact
that every trajectory contributes independently to our estimator,
meaning that the implementation is trivially parallelizable.

\section{Variance estimation in the small $\gamma$ limit}
\label{sec:vardos}

We know from~\eqref{eq:3} that the variance of our estimator is lower
than that of the direct estimator. In the specific context of a DOS
calculation using~\eqref{eq:langevin}, we can analyze the variance
more explicitly in the limit of small~$\gamma$, in which the descent
dynamics in~\eqref{eq:langevin} reduces to a closed equation for the
energy $E = \mathcal{H}(\qb,\pb)$ on the rescaled time $t'=\gamma t$.
This dynamics evolves on the disconnectivity (or Reeb)
graph~\cite{wales_archetypal_1998}, which branches at every energy
level at which a basin where $ \mathcal{H}(\qb,\pb)\le E$ splits into
more than one connected component.  In the simplest case when the
potential $U(\qb)$ has a single well, the graph has only one branch
and the value of $\gamma(\tau^E(\xb_i) - \tau^-(\xb_i))$ becomes the
same along every trajectory when $\gamma\to0$.  Therefore the
estimator~\eqref{eq:dos} has zero variance---a single trajectory gives
the exact value for $V(E)/V(E_{\text{max}})$.  If the disconnectivity
graph has several branches, we can count all the paths along the graph
starting at $E=E_{\text{max}}$ which end at a given branch.  Assuming
that the number of such paths is $M\ge1$, we can associate a
deterministic time $\Delta \tau^E_j>0$, possibly infinite, along each
path.  We define $\Delta \tau^E_j$ with $j=1,\ldots, M$ as the total
rescaled time the trajectory takes to go from
$\mathcal{H}(\qb,\pb)= E_{\text{max}}$ to $\mathcal{H}(\qb,\pb)=E$ by
the effective dynamics for $E$ along the path with index $j$ and
$\Delta \tau^E_j=\infty$ if the path terminates at an energy $E'>E$.
For any initial condition,
$\lim_{\gamma\to0} \gamma(\tau^E(\xb_i)- \tau^-(\xb_i) )=\Delta
\tau^E_j$ for some index $j$, meaning the only random component in the
procedure is which path is picked if the trajectory starts at $\xb_i$.
We denote by $p_j$ the probability, computed over all initial
conditions drawn uniformly in $\{\mathcal{H}(\qb,\pb) < E_{\max}\}$,
that the path with index $j$ is taken. Then in the small $\gamma$
limit the mean and variance of the estimator~\eqref{eq:dos} are
\begin{equation}
  \label{eq:2}
  \begin{aligned}
  \text{mean} &= \frac{V(E)}{V(E_{\text{max}})} = \sum_{j=1}^M p_j e^{-d \Delta
    \tau^E_j}, \\
  \text{variance} &= \sum_{j=1}^M p_j e^{-2d \Delta
    \tau^E_j} - \text{mean}^2.
  \end{aligned}
\end{equation}
The specific values of $p_j$, $\Delta\tau_j^E$, and $M$ which
determine the quality of the estimator~\eqref{eq:dos} depend on both
the structure of the disconnectivity graph and the effective equation
for the energy on this graph.  What is remarkable, however, is that
$p_j$, $\Delta\tau_j^E$, and $M$ depend on the dimensionality of the
system only indirectly.  In high dimensional settings, the complexity
of the disconnectivity is a generic challenge, but our approach has
favorable properties even in these difficult cases.  In particular,
the computational cost of the procedure increases only linearly in
$\gamma$ as we decrease this parameter to small values.  We also
stress that the formulae~\eqref{eq:2} for the mean and variance rely
on the assumption $\gamma\ll 1$, but the estimator remains valid for
any value of $\gamma.$

\begin{figure}
\includegraphics[width=.8\linewidth]{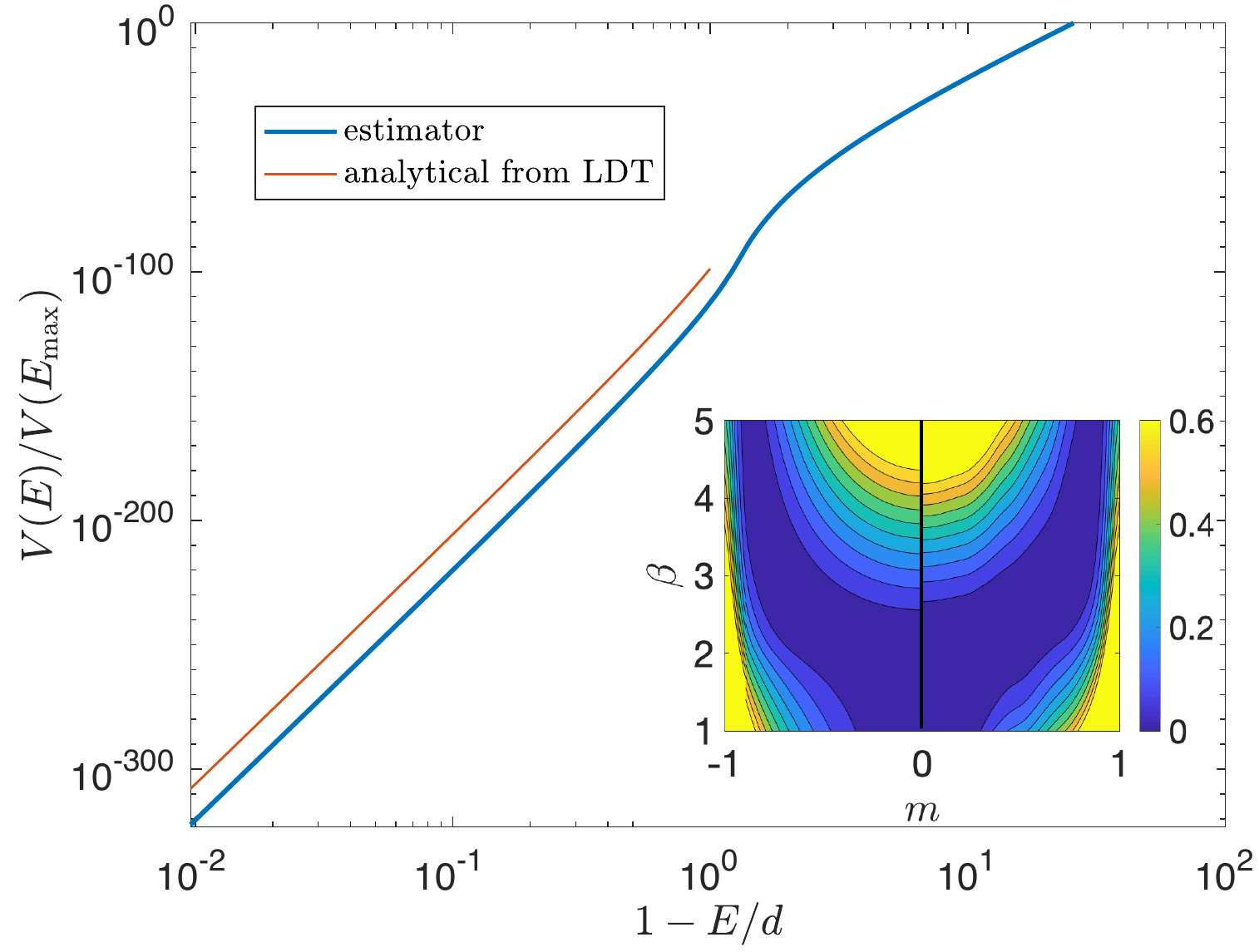}
\caption{Mean-field Ising model with potential~\eqref{eq:4}: volume
  ratio obtained from~\eqref{eq:dos} using a single ascent / descent
  trajectory. The inset shows the free energy in $\beta$ and
  $m = d^{-1} \sum_{i=1}^d \cos(q_i)$ that can also be estimated from
  this single trajectory (right half) and analytically using large
  deviation theory (left half).}
  \label{fig:simple_tests}
\end{figure}

\section{Phase diagram of the mean-field Ising model.}
\label{sec:ising}

As an illustration of the
statistical power contained in the nonequilibrium trajectories, we
computed the phase diagram of the mean-field Ising model with potential
\begin{equation}
  \label{eq:4}
  U(\qb) = -\frac1{2d} \sum_{i,j=1}^d \cos(q_i) \cos(q_j).
\end{equation}
This system is known to display a phase transition in temperature at
the critical $\beta_c=2$~\cite{martinsson_simulated_2018}. The
potential~\eqref{eq:4} is double-well, but because of the symmetry
$\qb\to-\qb$ a single ascent / descent trajectory can be used to
estimate the volume ratio $V(E)/V(E_{\text{max}})$.  We note that
while there are only to energy minima, the energy landscape has an
exponential number of critical points (the derivative with respect to
$\theta_j$ vanishes $2^d$ when $\sin(\theta_j)=0$), so the geometry of
the landscape is nontrivial.  We performed this calculation when
$d=100$ with $\gamma = 10^{-3}$ to obtain the result shown in
Fig.~\ref{fig:simple_tests}: as can be seen, this result spans 300
order of magnitudes and compares very well with the analytical
estimate that can be obtained in the large $d$ limit, as described in
Appendix~\ref{sec:meanfield Ising}. The single ascent /descent
trajectory can also be used to calculate the free energy in $\beta$
and magnetization $m = d^{-1} \sum_{i=1}^d \cos(q_i)$ of the system
(see Appendix~\ref{sec:meanfield Ising} for details).  The result
shown in the inset of Fig.~\ref{fig:simple_tests} demonstrates that
our estimate compares well with the large $d$ estimate.  The code to
reproduce these experiments is available on Gitlab~\footnote{Source
  code available at:
  \url{https://gitlab.com/rotskoff/trajectory_estimators}}.  The
numerical experiments require only several minutes of computation on a
single core, but parallelization strategies could dramatically reduce
the duration.

\section{Bayes factors}
\label{sec:bayesf}

The computations for the density of states have
an equivalent manifestation in Bayesian estimation.  Given a model
$\mathcal{M}$, one seeks to maximize the probability of a set of
parameters $\theta\in \mathbb{R}^d$ conditioned on observations of
data $\mathcal{D}.$ Using Bayes Theorem, we can write
\begin{equation}
  \mathbb{P}(\boldsymbol{\theta}| \mathcal{D}, \mathcal{M})
  = % \frac{P(\mathcal{D}| \boldsymbol{\theta}, \mathcal{M})
  %   P(\boldsymbol{\theta}|\mathcal{M})}{ P(\mathcal{D}|\mathcal{M})}
  % \equiv
  \mathcal{L}(\boldsymbol{\theta}) \pi(\boldsymbol{\theta})/Z
\end{equation}
where
$\mathcal{L}(\boldsymbol{\theta})=\mathbb{P} (\mathcal{D}|
\boldsymbol{\theta}, \mathcal{M}) $ is the likelihood function,
$\pi(\boldsymbol{\theta})=\mathbb{P}
(\boldsymbol{\theta}|\mathcal{M})$ is the prior, and
$Z=\mathbb{P} (\mathcal{D}|\mathcal{M}) = \int
\mathcal{L}(\boldsymbol{\theta}) \pi(\boldsymbol{\theta}) d
\boldsymbol{\theta}$ is the partition function, often called the
Bayesian evidence in this context; it is the canonical partition
function with $\beta=1.$

In Bayesian inference, we choose a model and then estimate its
parameters without knowledge of the partition function by doing
gradient descent on $-\log \mathcal{L}(\thetab)\equiv U(\thetab),$
which depends on the model we have taken.  However, there is no
\textit{a priori} guarantee that the chosen model is optimal, so it is
often necessary to make comparisons of two distinct models
$\mathcal{M}$ and $\mathcal{M'}$.  Ideally, one would compare the
probability of the observed data given each model, that is the Bayes
factors
\begin{equation}
  Z/ Z' = \mathbb{P} (\mathcal{D}|\mathcal{M})/\mathbb{P}
  (\mathcal{D}|\mathcal{M}').
\end{equation}
Similarly, computing posterior probabilities also requires knowledge
of the partition function.

As we have already emphasized, computing $Z$ is intractable
analytically in all but the simplest cases.
\cite{skilling_nested_2006} demonstrated that it is possible to
numerically evaluate the ``prior volume'',
\begin{equation}
  V(L) = \int_{\mathcal{L}(\boldsymbol{\theta})\ge L }
  \pi(\boldsymbol{\theta}) d\boldsymbol{\theta}
\end{equation}
to produce an estimate of $Z$ via
\begin{equation}
Z = \int_0^{L_0} V(L) \ dL,
\label{eq:partition}
\end{equation}
where $L_0$ is the maximum value of the likelihood.  Just
as in the density of states calculation, we can evaluate the Bayesian
evidence by using trajectorial estimators.

To do so, we sample parameters of the model $\mathcal{M}$ uniformly
and define a flow of parameters via dissipative Langevin dynamics with
$ U(\thetab) = -\log \mathcal{L}(\thetab)$ (which also gives $L_0$
from the terminal point of the descent trajectories).  We construct an
estimate of $Z$ by computing $V(L)$ using Eq.~\eqref{eq:dos} and
numerically integrating Eq.~\eqref{eq:partition} using quadrature.
Note that the contribution from the momenta can be factored out and
the resulting Gaussian integral can be computed exactly.

\begin{figure}
\includegraphics[width=.8\linewidth]{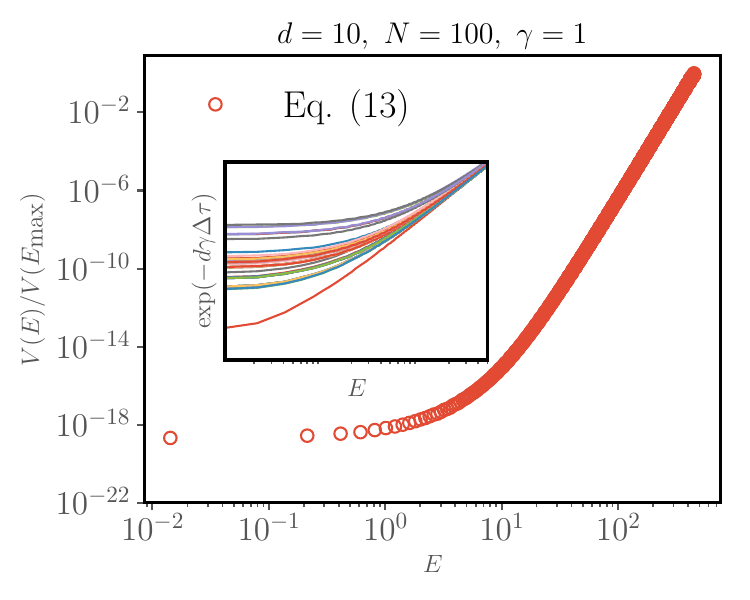}
%\end{center}
%
\caption{Mixture of Gaussians inference problem with $d=10$ and 50 wells in the mixture. The volume of states below $E=-\log{\mathcal{L}}$ is shown as red circles. In the small gamma limit, the time to reach energy $E$, $\Delta \tau = \tau^E - \tau^{-}$, should be independent of the initial condition. 
The inset shows that the dissipative dynamics converges to many different local minima, corresponding to the individually discernible lines in the inset. 
A total of $N=100$ trajectories are plotted, but there are fewer than one hundred visible lines, meaning that some trajectories not only end in the same basin, but also have decay in energy at the same rate, as the small $\gamma$ limit predicts. 
This demonstrates that the low variance regime can be achieved even with modest values of  $\gamma.$}
\label{fig:bayes}
\end{figure}

We tested our approach using a mixture of Gaussians model, a benchmark
which has been used to characterize nested sampling for inference
problems~\cite{feroz_multimodal_2008}.  The model is defined as a
mixture of $n$ distributions in dimension $d$ with amplitudes $A_i$,
\begin{equation}
  \mathcal{L}(\thetab) = \sum_{i=1}^n
  A_i e^{ -\tfrac12(\thetab-\boldsymbol{\mu}_i)^T\boldsymbol{\Sigma}_i^{-1} (\thetab-\boldsymbol{\mu}_i)}.
  \label{eq:mog}
\end{equation}
Though we do not have access to the exact expression for $V(E)$ at all
energy levels in this model, we can evaluate the partition function
$Z$ exactly.

We used $n=50$ wells with depths exponentially distributed in
dimension $d=10$, an example much more complex than previous
benchmarks. While this landscape is not rugged, in mixture of Gaussian
problems entropic effects can lead to extremely difficult optimization
problems because we are required to sample exponentially small
volumes.  In this regime, brute force Monte Carlo approaches fail
dramatically.  Fig.~\ref{fig:bayes} illustrates the statistical power
of the trajectory reweighting approach. With only 100 trajectories, we
recover the volume of states for the likelihood function extremely
accurately, especially at low energies, where the standard error is
vanishingly small.  An accurate estimate at low energies leads to
robust estimates of $Z$ because the contribution to $Z$ decays
exponentially with $E$. In particular, we know the low energy volume
estimates are accurate because we compute $Z=17.41$ versus the exact
result $Z=17.10$. For this calculation we set $E_\text{max}= 450$,
meaning that the states we neglected have likelihood lower than
$e^{-450}$.

\section{Conclusions}
\label{sec:conclu}

Any estimate of the microcanonical partition function requires a
thorough exploration of the states of the system.  Both naive Monte
Carlo sampling and equilibrium dynamics often fail to visit states,
which, though rare, dramatically impact the thermodynamic properties
of the system.  A nonequilibrium dynamics suffers from precisely the
opposite problem: it explores the states rapidly, but not in
proportion to their equilibrium probabilities.  Our estimator, via
Eq.~\eqref{eq:dos} establishes a simple link between a nonequilibrium
dynamical observable and a static property, the volume of phase space.

With a properly formulated algorithm, we can fully account for the
statistical bias of a nonequilibrium dynamics.  The resulting
estimators can access states that are extremely atypical in
equilibrium sampling schemes, but nevertheless physically
consequential.  While we demonstrated the potential of these
estimators by computing the density of states and the computationally
analogous Bayes factor, the expression in~\eqref{eq:estimator} is
extremely general.  Attractive applications within reach include
adapting this approach to basin volume
calculations~\cite{asenjo_numerical_2014,martiniani_turning_2016},
computing the partition function of restricted Boltzmann
machines~\cite{neal_annealed_2001,NIPS2013_4879}, and importance
sampling to compute properties of systems in nonequilibrium stationary
states, like active
matter~\cite{bechinger_active_2016,cates_motility-induced_2015}.

\section*{Acknowledgments}
The authors thank Daan Frenkel, Stefano Martiniani, K. Julian Schrenk,
and Shang-Wei Ye for discussions that helped motivate this
work. G.M.R. acknowledges support from the James S. McDonnell
Foundation. E.V.E. was supported by National Science Foundation (NSF)
Materials Research Science and Engineering Center Program Award
DMR-1420073; and by NSF Award DMS-1522767.

\bibliographystyle{abbrv}
\bibliography{refs-final}

\appendix

\section{Derivation of expression \eqref{eq:rrhoneqexplicit} for
  $\rho_{\text{ne}}(\xb)$ }
\label{sec:derirhone}

To derive the expression in~\eqref{eq:rrhoneqexplicit} for the
density $\rho_{\text{ne}}(\xb)$, consider first the forward time
trajectories alone and let us define the forward density
$\rho_{\text{ne}}^+(\xb)$ via
\begin{equation}
  \label{eq:rrhoneqplusAA}
  \begin{aligned}
    \avg{\phi}^+_{{\rm ne}} &= \frac1{\avg{\tau^+} } \int_{\Omega}
    \int_{0}^{\tau^+(\xb)} \phi(\Xb(t,\xb)) dt \, \rho(\xb) d\xb\\
    & \equiv \int_{\Omega}\phi(\xb) \rho^+_{\text{ne}}(\xb) d\xb
  \end{aligned}
\end{equation}
The nonequilibrium density $\rho_{\text{ne}}^+ (\xb) $ satisfies the
stationary Liouville equation
\begin{equation}
  \avg{\tau^+}^{-1} \rho(\xb) = \div \left( \bb(\xb) \rho^+_{\rm ne}(\xb) \right),
  \label{eq:liouville}
\end{equation}
with the boundary condition $\rho^+_{\rm ne}(\xb) =0$ on the regions
of $\partial \Omega$ where $\bb(\xb)$ points inward (since, by
construction, no mass can be transported there forward in time).
Physically, this equation asserts that, at stationarity, the
nonequilibrium probability flux out of a small volume in the vicinity
of $\xb$ is balanced by the rate of reinjection.  To
solve~\eqref{eq:liouville} notice that if we differentiate
$\rho_{\rm ne}^+(\Xb(t,\xb))$ with respect to time, by the chain rule
we have
\begin{equation}
  \label{eq:liou1}
  \begin{aligned}
    \d{}{t} \rho_{\rm ne}^+ = \bb\cdot \grad \rho_{\rm ne}^+=
    \avg{\tau^+}^{-1}\rho- (\div \bb) \rho_{\rm ne}^+
  \end{aligned}
\end{equation}
where all functions are evaluated at $\Xb(t, \xb) $ and we used
Eq.~(1) for $\Xb(t, \xb) $ to derive the first equality
and~\eqref{eq:liouville} to derive the second.  Using the
Jacobian (Eq.~(5)),
\begin{equation}
  J(t,\xb) = \exp \left({\textstyle\int_0^t \div \bb (\Xb(s, \xb))
      ds} \right),
\end{equation}
we can write~\eqref{eq:liou1} as
\begin{equation}
  \begin{aligned}
    \d{}{t} \left(\rho_{\rm ne}^+(\Xb(t, \xb))J(t,\xb) \right) =
    \avg{\tau^+}^{-1}\rho (\Xb(t, \xb)) J(t,\xb).
  \end{aligned}
\end{equation}
Integration from $t=\tau^-(\xb)$ to $t=0$ using the boundary condition
$\rho_{\rm ne}^+(\Xb(\tau^-(\xb), \xb)) =0$ gives
\begin{equation}
  \label{eq:rhoneqplus}
  \rho_{\rm ne}^+(\xb)
  = \avg{\tau^+}^{-1} \int_{\tau^-(\xb)}^0J(t,\xb)
  \rho(\Xb(t,\xb)) dt.
\end{equation}
A similar calculation gives the nonequilbrium stationary density resulting from
the reverse time propagation of the dynamics, $\rho_{\rm ne}^-(\xb)$,
defined via
\begin{equation}
  \label{eq:rrhoneqplus}
  \begin{aligned}
    \avg{\phi}^-_{\rm ne} &=\frac1{\avg{\tau^-} } \int_{\Omega}
    \int^{0}_{\tau^-(\xb)} \phi(\Xb(t,\xb)) dt \,
    \rho(\xb) d\xb\\
    & \equiv \int_{\Omega}\phi(\xb) \rho^-_{\text{ne}}(\xb) d\xb
  \end{aligned}
\end{equation}
We obtain
\begin{equation}
  \label{eq:rhoneqminus}
  \rho_{\rm ne}^-(\xb)
  = \avg{\tau^-}^{-1} \int^{\tau^+(\xb)}_0J(t,\xb)
  \rho(\Xb(t,\xb)) dt.
\end{equation}
The nonequilibrium density defined via Eq.~(4) can then be expressed
by superposing $\rho_{\rm ne}^+(\xb)$ and $\rho_{\rm ne}^-(\xb)$:
\begin{equation}
  \label{eq:rrhoneqexplicitA}
  \begin{aligned}
    \rho_{\textrm{ne}}(\xb) & = \frac{\avg{\tau^+}\rho_{\rm ne}^+(\xb)
      - \avg{\tau^-}\rho_{\rm ne}^-(\xb)}{\avg{\tau}}
    \\
    & = \avg{\tau}^{-1}
    \int_{\tau^-(\xb)}^{\tau^{+}(\xb)}J(t,\xb)\rho(\Xb(t,\xb))
    dt
  \end{aligned}
\end{equation}
This is~\eqref{eq:rrhoneqexplicit}

\section{Derivation of expression~\eqref{eq:estimator0} for the
  estimator}
\label{Sec:derivest1}
To derive the estimator in~\eqref{eq:estimator0}, start from
\eqref{eq:reweight} and write
\begin{equation}
  \begin{aligned}
    \avg{\phi} & = \avg{ \phi \rho/ \rho_{\text{ne}}}_{\text{ne}}\\
    & = \frac1{\avg{\tau}}\int_{\mathbb{R}^d}
    \int_{\tau^-(\xb)}^{\tau^+(\xb)} \frac{\phi(\Xb(t,\xb))
      \rho(\Xb(t,\xb))}{\rho_{\text{ne}}(\Xb(t,\xb))} dt \, \rho(\xb)
    d\xb\\
    & = \int_{\mathbb{R}^d}
    \int_{\tau^-(\xb)}^{\tau^+(\xb)}\frac{\phi(\Xb(t,\xb))\rho(\Xb(t,\xb))
      }{\int_{\tau^-(\Xb(t,\xb))}^{\tau^+(\Xb(t,\xb))}J(s,
      \Xb(t,\xb))
      \rho(\Xb(s, \Xb(t,\xb)))
      ds} dt \, \rho(\xb) d\xb.
  \end{aligned}
  \label{eq:estimator0A}
\end{equation}
Using $\Xb(s,\Xb(t,\xb)) = \Xb(s+t,\xb)$ we have
\begin{equation}
  \label{eq:1AA}
  \begin{aligned}
    J(s,
      \Xb(t,\xb)) & = \exp \left({\textstyle\int_0^s \div \bb (\Xb(u, \Xb(t,\xb)))
          du} \right)\\
      & = \exp \left({\textstyle\int_0^s \div \bb (\Xb(u+t,\xb))
          du} \right)
      \\
      & = \exp \left({\textstyle\int_t^{s+t} \div \bb (\Xb(u',\xb))
          du'} \right)\\
      & = \frac{J(s+t,\xb)}{J(t,\xb)}
  \end{aligned}
\end{equation}
which implies that we can transform the denominator
in~\eqref{eq:estimator0} into
\begin{equation}
  \label{eq:2AA}
  \begin{aligned}
    & \int_{\tau^-(\Xb(t,\xb))}^{\tau^+(\Xb(t,\xb))}J(s,
    \Xb(t,\xb)) \rho(\Xb(s, \Xb(t,\xb)))
    ds\\
    =
    &\frac1{J(t,\xb)}\int_{\tau^-(\Xb(t,\xb))}^{\tau^+(\Xb(t,\xb))}
    J(s+t,\xb)
    \rho(\Xb(s+t,\xb))) ds
    \\
    =
    &\frac1{J(t,\xb)}\int_{\tau^-(\Xb(t,\xb))+t}^{\tau^+(\Xb(t,\xb))+t}
    J(s',\xb)
    \rho(\Xb(s',\xb))) ds'
    \\
    =
    &\frac1{J(t,\xb)}\int_{\tau^-(\xb)}^{\tau^+(\xb)}
    J(s',\xb)
    \rho(\Xb(s',\xb))) ds'
  \end{aligned}
\end{equation}
Inserting this expression in~\eqref{eq:estimator0} gives
\begin{equation}
  \begin{aligned}
    \avg{\phi} &= \int_{\mathbb{R}^d}
    \int_{\tau^-(\xb)}^{\tau^+(\xb)}\frac{\phi(\Xb(t,\xb))J(t,\xb)\rho(\Xb(t,\xb))
    }{\int_{\tau^-(\xb)}^{\tau^+(\xb)}J(s',\xb)\rho(\Xb(s',\xb))
      ds'} dt \, \rho(\xb) d\xb\\
    & = \int_{\mathbb{R}^d}
    \frac{\int_{\tau^-(\xb)}^{\tau^+(\xb)}\phi(\Xb(t,\xb))J(t,\xb)\rho(\Xb(t,\xb))
      dt}{\int_{\tau^-(\xb)}^{\tau^+(\xb)}J(t,\xb)\rho(\Xb(t,\xb))
      dt} \, \rho(\xb) d\xb
  \end{aligned}
  \label{eq:estimator00}
\end{equation}
which is~\eqref{eq:estimator0}. Note that this equality holds even if
$\tau_-(\xb) = -\infty$ and/or $\tau_+(\xb) = +\infty$, as long as the
time integrals converge. That is, we don't necessarily need to
introduce target sets to make the descent / ascent trajectories
finite.

\section{Fixed time equivalent of \eqref{eq:estimator0}}
\label{sec:fixedest}

Suppose that $\Xb(t,\xb)\in \Omega$ for all $\xb\in \Omega$ and $t\in
\RR$. Then, given any $-\infty\le T_1< T_2 \le \infty$, we have the following
estimator
\begin{equation}
  \label{eq:4AA}
  \<\phi \> = \left\<
  \int_{T_-}^{T_+}  \frac{\phi(\Xb(t)) 
    \rho(\Xb(t)) J(t) }
  {\int_ {t-T_+}^{t-T_-}
    \rho(\Xb(s)) J(s) ds } dt\right\>
\end{equation}
For $T_-=-\infty$ and $T_+=\infty$, assuming that the integrals
converge, this estimator can also be written as
\begin{equation}
  \label{eq:12AA}
  \< \phi\> = \left\<
  \frac{\int_{-\infty}^\infty \phi(\Xb(t)) 
    \rho(\Xb(t)) J(t)dt}
  {\int_{-\infty}^\infty 
    \rho(\Xb(t)) J(t)  dt}\right\>
\end{equation}
This is the same as \eqref{eq:estimator0} when $\tau_-(\xb) =\-infty$
and $\tau_+(\xb)=+\infty$ for all $\xb\in \Omega$.

To establish~\eqref{eq:4AA}, notice that its  left hand side is 
\begin{equation}
    \label{eq:estimator1steps}
    \begin{aligned} 
      \text{LHS\eqref{eq:4AA}}
      & = \int_{\RR^{n}} \int_{T_-}^{T_+} \frac{\phi(\Xb(t,\xb)) \rho(\Xb(t,\xb)) J(t,\xb)}
      {\int_{t-T_+}^{t-T_-}
        \rho(\Xb(s,\xb)) J(s,\xb)ds} dt \rho(\xb) d\xb\\
      & = \int_{T_-}^{T_+}  \int_{\RR^{n}} \frac{\phi(\Xb(t,\xb)) \rho(\Xb(t,\xb)) J(t,\xb)}
      {\int_{t-T_+}^{t-T_-}
        \rho(\Xb(s,\xb)) J(s,\xb)ds} \rho(\xb) d\xb dt
   \end{aligned}
 \end{equation}
 Now use $\xb'= \Xb(t,\xb)$ as new integration variable, the group
 property of the flow to set $\xb = \Xb(-t,\xb')$ and
 $\Xb(s,\xb)=\Xb(s,\Xb(-t,\xb'))=\Xb(s-t,\xb')$, as well as
 $J(s,\Xb(-t,\xb')) = J(s-t,\xb')/J(-t,\xb')$ to
obtain
 \begin{equation}
    \label{eq:estimator1stepsb}
    \begin{aligned} 
      \text{LHS\eqref{eq:4AA}}
      & = \int_{T_-}^{T_+} \int_{\RR^{n}} \frac{
        \phi(\xb') \rho(\xb') /J(-t,\xb') 
      }{\int_{t-T_+}^{t-T_-} \rho(\Xb(s-t,\xb'))
        J(s-t,\xb')/J(-t,\xb')  ds}  \rho(\Xb(-t,\xb')) J(-t,\xb') d\xb'
      dt\\
      & = \int_{T_-}^{T_+} \int_{\RR^{n}} \frac{
        \phi(\xb') \rho(\xb') 
      }{\int_{t-T_+}^{t-T_-} \rho(\Xb(s-t,\xb'))
        J(s-t,\xb') ds}  \rho(\Xb(-t,\xb'))  J(-t,\xb')  d\xb' dt
       \end{aligned}
 \end{equation}
 Finally use $s'=t-s$ as new integration variable in the integral at the
 denominator to deduce
 \begin{equation}
    \label{eq:estimator1stepsc}
    \begin{aligned} 
      \text{LHS\eqref{eq:4AA}}
      & =  \int_{T_-}^{T_+} \int_{\RR^{n}} \frac{
        \phi(\xb') \rho(\xb') 
      }{\int_{T_-}^{T_+} \rho(\Xb(-s',\xb'))
        J(-s',\xb') ds'}  \rho(X(-t,\xb'))  J(-t,\xb')  d\xb' dt
      \\
      & = \int_{\RR^{n}} \phi(\xb') \rho(\xb') \frac{
        \int_{T_-}^{T_+} \rho(\Xb(-t,\xb'))  J(-t,\xb')dt      
      }{\int_{T_-}^{T_+} \rho(\Xb(-s',\xb'))
        J(-s',\xb') ds'}   d\xb' \\
      & = \int_{\RR^{n}} \phi(\xb') \rho(\xb') d\xb'  \equiv \<\phi\>.
    \end{aligned}
\end{equation}

\section{Eq.~\eqref{eq:estimator0} is a zero-variance estimator in
  one-dimension}
\label{sec:zerovar}

Suppose that $x\in \RR$, let $b(x) = 1$ so that $X(t,x) = x + t$ and
$J(t,x) = 1$ , and consider the one-dimensional version
of~\eqref{eq:12AA}
\begin{equation}
  \label{eq:10}
  \<\phi \> = \left\<
  \frac{\int_{-\infty}^\infty \phi(X(t)) 
    \rho(X(t)) dt}
  {\int_{-\infty}^\infty 
    \rho(X(t))  dt}\right\>
\end{equation}
The factor under the expectation at the left hand side is
\begin{equation}
  \label{eq:11}
  \frac{\int_{-\infty}^\infty \phi(X(t,x)) 
    \rho(X(t,x)) dt}
  {\int_{-\infty}^\infty 
    \rho(X(t,x))  dt}  = \frac{\int_{-\infty}^\infty \phi(x+t) 
    \rho(x+t) dt}
  {\int_{-\infty}^\infty 
    \rho(x+t)  dt} = \<\phi\>
\end{equation}
where we used $x'=x+t$ as new integration variable to get the second
equality. Since \eqref{eq:11} hold for any $x\in\RR$, the expectation
in~\eqref{eq:10} is unnecessary, i.e. \eqref{eq:11} is a zero variance
estimator of $\<\phi\>$. It is easy to see that this remains true as long as the
velocity $b(x)$ is stays bounded away from zero (i.e. does not change
sign), so that $X(t,x) \to -\infty $ as $t\to -\infty$ and
$X(t,x) \to \infty $ as $t\to \infty$ if $b(x)>0$ and conversely if
$b(x)<0$.

Of course, the one-dimensional situation is very special since in that
case we can trade space for time. In higher dimension, this is no
longer posible, and the estimator~\eqref{eq:12AA} will not have zero
variance in general. Still, it suggests that by carefully designing the
velocity field $\bb(\xb)$ it may be possible to get low or even zero
variance estimators using~\eqref{eq:12AA}.

\section{Comparison with Neal's Annealed Importance Sampling (AIS)
  method}
\label{sec:AIScomp}

While our method is conceptually similar to AIS, the estimator (13)
has important philosophical and practical differences from AIS, which
we highlight here.

In AIS, one defines a sequence of distributions in order to estimate
an expectation with respect to a target density $\rho_1$, usually
known only up to a normalization factor, i.e. we know some
$\hat \rho_1(\xb)$ can be evaluated pointwise, but the factor
$Z_1 = \int_\Omega \hat \rho_1(\xb) d\xb$ needed to get the normalized
density $\rho_1(\xb) = Z_1^{-1} \hat \rho_1(\xb)$ is not known. In
practice, the sampling is done by transporting from an initial density
$\rho$ using a nonequilibrium dynamics: $\rho$ can be sampled
e.g. using Metropolis-Hastings Monte-Carlo, and is also only known up
to a normalization factor in general. This transport defines a
nonequilibrium density $\rho_{\text{ne}}(\xb)$, also known only up to
a normalization factor, so that the annealed importance sampling
scheme for the expectation of an observable $\phi$ is
\begin{equation}
  \label{eq:AISestimator}
  \langle{\phi}\rangle_{\rho_1} =
  \frac{\langle \phi \hat \rho_1/\hat \rho_{\rm ne}\rangle_{\rm ne}}
  {\langle\hat \rho_1/\hat \rho_{\rm ne}\rangle_{\rm ne}},
\end{equation}
where the ratio $\hat \rho_1/\hat \rho_{\rm ne}$ plays the role of the
weights in AIS.

The situation with our estimator is quite different. First, we use
$\rho_1=\rho$. Secondly, we generate the trajectories in a way such
that, even if we do not know the normalization of $\rho_1=\rho$, we
have the property that
\begin{equation}
  \langle\hat \rho_1/\hat \rho_{\rm ne}\rangle_{\rm ne}=
  \langle\hat \rho/\hat \rho_{\rm ne}\rangle_{\rm ne} = 1
\end{equation}
by construction.  Therefore our reweighting scheme does not require
that we estimate a ratio of expectations; therefore, it provides us with an
unbiased estimator, unlike~\eqref{eq:AISestimator}. Finally, this
estimator has lower variance than the direct sample mean estimator, as shown by Eq.~(9)
in the main text.

\section{The mean-field Ising model}
\label{sec:meanfield Ising}

We next consider a continuous
version of the Curie-Weiss magnet, i.e. the mean-field Ising model
with $d$ spins and potential~\eqref{eq:4}
\begin{equation}
  \label{eq:CWpot}
  U(\qb) = - \frac{1}{2d} \sum_{i,j=1}^{d} \cos q_i \cos q_j
  = - \frac{1}{2d} \left( \sum_{i=1}^{d} \cos q_i \right)^2
\end{equation}
The Gibbs (canonical) density for this model is
\begin{equation}
  \label{eq:gibbstheta}
  \rho_c(\qb) =
  Z^{-1}(\beta) e^{-\beta U(\qb)}\quad\text{where} \quad 
  Z(\beta) = 
  \int_{[-\pi,\pi]^{d}} e^{-\beta U(\qb)} d\qb.
\end{equation}
This system has similar thermodynamics properties as the standard
Curie-Weiss magnet with discrete spins, but it is amenable to
simulation by Langevin dynamics since the angles $q_i$ vary
continuously.

In particular, like the standard Curie-Weiss magnet, the system with
potential~\eqref{eq:CWpot} displays phase-transitions when $\beta$ is
varied. To see why, and also introduce the scaled free energy that we
monitor in our numerical experiments, let us marginalize the Gibbs
density~\eqref{eq:gibbstheta} on the average magnetization $m$ defined
as
\begin{equation}
  \label{eq:19}
  m = \frac1d\sum_{i=1}^{d} \cos q_i.
\end{equation}
This marginalized density is given by
\begin{equation}
  \label{eq:25}
  \bar \rho_c(m)  = \int_{[-\pi,\pi]^{K}} \rho_c(\qb)
  \delta \left( m -\frac1d\sum_{i=1}^{d} \cos q_i
  \right)
  d\qb .
\end{equation}
A simple calculation shows that
\begin{equation}
  \label{eq:gibbsm}
  \bar \rho_c(m) = \bar Z^{-1}(\beta) e^{-\beta d F_d(m,\beta)}
  \quad\text{where}\quad
  \bar Z(\beta) = \int_{-1}^{1} e^{-\beta d F_d(m,\beta)} dm
\end{equation}
Here
we introduced the (scaled) free energy $F_d(m,\beta)$ defined as
\begin{equation}
  \label{eq:33}
  F_d(m,\beta) = V(m)- \beta^{-1} S_d(m)
\end{equation}
with potential term
\begin{equation}
  \label{eq:CWpot2}
  V(m) = -\tfrac12 m^2 
\end{equation}
and entropic term 
\begin{equation}
  \label{eq:cwent}
  S_d(m) = d^{-1} \log 
  \int_{[-\pi,\pi]^{d}} \delta \left( m - \frac1d\sum_{i=1}^{d} \cos q_i
  \right)
  d\qb .
\end{equation}
The marginalized density~\eqref{eq:gibbsm} and the free
energy~\eqref{eq:33} can be used to analyze the properties of the
system in thermodynamic limit when $d\to\infty$ and map out its phase
transition diagram in this limit. In particular, standard results from
large deviation theory recalled below can be used to show that
$F_d(m,\beta)$ has a limit as $d\to\infty$ that has a single minimum
at high temperature, but two minima at low temperature. Since
$F_d(m,\beta)$ is scaled by $d$ in~\eqref{eq:gibbsm}, this implies
that density can become bimodal at low temperature, indicative of the
presence of two strongly metastable states separated by a free energy
barrier whose height is proportional to $d$.

The limiting free energy $F(m,\beta)$ is defined as
\begin{equation}
  F(m,\beta) = \lim_{d \to \infty}F_d(m) =
  -\tfrac12 m^2 -\beta^{-1} \lim_{d \to \infty}S_d(m).
\label{eq:FreeEnergy}
\end{equation}
To calculate the limit of the entropic term, let us define
$H(\lambda)$ via the Laplace transform of \eqref{eq:cwent} through
\begin{equation}
\begin{split}
  e^{ -d H(\lambda)} & = \int_{-1}^1 e^{-d \lambda m + d S_d(m)} dm\\
  &= \int_{[-\pi,\pi]^d} e^{-\lambda \sum_{i=1}^{d} \cos(q_i)}
  d\qb
  \\
  &= \prod_{i=1}^{d} \int_{-\pi}^{\pi} e^{-\lambda \cos (q_i)} 
  dq_i\\
  &= \left( 2 \pi I_0(\lambda) \right)^d,
\end{split}
\end{equation}
where $I_0(\lambda)$ is a modified Bessel function.  In the large $d$
limit, $S(m)$ can be calculated from $H(\lambda)$ by Legendre
transform
\begin{equation}
\begin{split}
  S(m) = \lim_{d\to\infty} S_d(m) &= \min_{\lambda} \left\{ \lambda m
    - H(\lambda) \right\} \\
  &= \min_{\lambda} \left\{ \lambda m + \log I_0(\lambda) \right\} +
  \log (2\pi).
\label{eq:105}
\end{split}
\end{equation}
The minimizer $\lambda(m)$ of~\eqref{eq:105} satisfies
\begin{equation}
  \label{eq:26}
  m  =  -\frac{I_1(\lambda(m))}{I_0(\lambda(m))}
\end{equation}
which, upon inversion, offers a way to parametrically represent $S(m)$
using
\begin{equation}
  \label{eq:27}
  \begin{aligned}
    S(m(\lambda)) = \lambda m(\lambda) + \log I_0(\lambda) + \log
    (2\pi), \qquad m(\lambda) = -\frac{I_1(\lambda)}{I_0(\lambda)},
    \qquad \lambda \in \mathbb{R}.
  \end{aligned}
\end{equation}
Similarly we can represent $F(m,\beta)$ as 
\begin{equation}
  \label{eq:28}
  F(m(\lambda),\beta)= -\tfrac12 m^2(\lambda) - b m(\lambda)
  -\beta^{-1} S(m(\lambda)), \qquad m(\lambda) =
  -\frac{I_1(\lambda)}{I_0(\lambda)} \qquad \lambda \in \RR.
\end{equation}
These are the formulae we use to plot the free energy shown in the
inset of Fig.~\ref{fig:simple_tests}. We compare this free energy with the one obtained
by estimating the following expectation using our estimator with a
single ascent / descent trajectory
\begin{equation}
  \label{eq:3AA}
  -(d\beta)^{-1} \log \left\langle e^{-\beta U(\qb)}
    \delta \left( m - \frac1d \sum_{i=1}^d \cos q_j\right)\right\rangle
\end{equation}

A similar calculation can be used to estimate the factor $Z(\beta)$
in~\eqref{eq:gibbstheta} using the identity
\begin{equation}
  \label{eq:4AA}
  Z(\beta) = \int_{-1}^1 e^{\frac12 \beta d m^2 + d S_d(m)} dm
\end{equation}
In the $d\to\infty$ limit this implies that
\begin{equation}
  \label{eq:5}
  G(\beta ) \equiv \lim_{d\to\infty} d^{-1} \log Z(\beta) = \max_{m\in[-1,1]}
  \left(\tfrac12 \beta d m^2 + d S(m)\right)
\end{equation}
Since the density of states
\begin{equation}
  \label{eq:7}
  D(E) = \int_{[-\pi,\pi]^d \times \RR^d}
  \delta \left(E - \frac12 |\pb|^2 -U(\qb)\right) d\qb d\pb
\end{equation}
is related to $Z(\beta)$ as (accounting for the extra factor coming
from the momenta)
\begin{equation}
  \label{eq:6}
  (2\pi \beta^{-1})^{d/2} Z(\beta) = \int_\RR D(E) e^{-\beta E} dE
\end{equation}
we have
\begin{equation}
  \label{eq:9}
   \lim_{d\to\infty} d^{-1} \log D(d \mathcal{E}) = Q(\mathcal {E})
\end{equation}
with
\begin{equation}
  \label{eq:8}
  Q(\mathcal {E})= - \min_{\beta} \left(\beta \mathcal{E}
    +G(\beta) - \tfrac12 \log \beta\right) +\tfrac12 \log\pi.
\end{equation}
This is the formula we used to plot the red curve in
Fig.~\ref{fig:simple_tests} used for comparison with the estimate of
$V(E)$ we obtained directly using our estimator.

\end{document}